# Electro-deposition and Mechanical Stability at Lithium/Solid Electrolyte Interface during Plating in Solid-State Batteries

Qingsong Tu [ab], Luis Barroso-Luque [a], Tan Shi [a], Gerbrand Ceder [ab*@]

a. Department of Materials Science and Engineering, University of California, Berkeley, CA 94720, USA

b. Materials Sciences Division, Lawrence Berkeley National Laboratory, Berkeley, CA 94720, USA

*Corresponding author Email: gceder@berkeley.edu

@ Lead contact

# Summary

Interfacial deposition stability between Li metal and a solid electrolyte (SE) is important in preventing interfacial contact loss, mechanical fracture, and dendrite growth in Li-metal solid state batteries (SSB). In this work, we investigate the deposition and mechanical stability at the Li metal/SE interface and its consequences (such as SE fracture and contact loss). A wide range of contributing factors are investigated, such as charge and mass transfer kinetics, plasticity of Li metal and fracture of the SE, and the applied stack pressure. We quantify the effect of the ionic conductivity of the SE, the exchange current density of the interfacial charge-transfer reaction and SE surface roughness on the Li deposition stability at the Li metal/SE interface. We also propose a "mechanical stability window" for the applied stack pressure that can prevent both contact loss and SE fracture, which can be extended to other metal-electrode (such as Sodium) SSB systems.

# Introduction

Li-metal all-solid-state batteries (SSBs) are promising candidates to replace conventional secondary batteries, providing improved safety with the use of non-flammable solid electrolytes (SE) [1,2], and higher energy density with the use of a Li-metal anode. [3,4] However, several bottlenecks originating from the intrinsic properties of Li metal and SEs hinder the practical application of Li-metal SSBs. First, although ceramic SEs have elastic moduli that are several times higher than that of Li metal, and are predicted to be stiff enough to suppress Li growth in the SE [5,6], recent experiments have revealed the presence of metallic Li in the SE, which can lead to cell shorting and potential safety issues. [7,8] Second, the high interfacial overpotential limits the power density and rate performance of Li-metal SSBs [9], which has been attributed to poor mechanical contact at the Li metal/SE interface [10,11] and the slow charge transfer kinetics stemming from the unstable interface between the SE and Li metal. [12,13] Substantial efforts have been devoted to design a system with stable Li deposition and intimate interfacial contact, including new design of porous electrodes to suppress dendrite growth [14,15], the use of artificial interlayers to eliminate SEI formation [16,17], and application of mechanical stress to reduce the interfacial resistance. [18-20]

Theoretical investigations are important for understanding the behavior at the Li metal/SE interface considering the difficulty in characterizing the buried interface between Li metal and the SE. Monroe and Newman investigated the role of mechanical stress in the growth stability of a planar interface [5], and suggested that a high shear modulus of the electrolyte (at least two times higher than that of Li metal) removes the inherent instability of planar growth, and by extension, may limit dendrites. [6] Their work has been further extended by considering the plasticity of Li metal [21] and the large shear modulus of SEs. [22] In their derivations, the interfacial stress was believed to be the main factor that affects the surface overpotential, which in turn changes the current density distribution at the Li metal/SE interface. Therefore, they concluded that mechanical properties of the Li metal and the SE (such as Shear moduli, Poisson's ratio) were dominant factors for the Li deposition stability. However, it has been demonstrated that many other factors, such as interfacial geometry [23], charge-transfer kinetics on the interface [24], ion transfer in bulk SE [25], and the externally applied compressive stress [26], can all influence the stability.

In general, uneven distributions of current and stress at the Li metal/SE interface (due to geometric irregularity [23] or material inhomogeneity [27]) lead to unstable Li electrodeposition (dendrite formation) and unstable morphological evolution (interfacial contact loss). Slow ionic transport and/or slow charge transfer between Li metal and SE may increase the area-specific resistance (ASR) of the cell [28,29] but the latter can also suppress dendrite formation by creating a more homogeneous current distribution. An externally applied compressive stress ("stack pressure") in the range of $1 \sim 10\ MPa$ when Sulfide SE are used in the cell [30,31]) may alleviate physical contact loss through deformation and flow of Li metal [26], but can also facilitate dendrite propagation when stress intensification at specific locations of the interface (such as the tips of defects) leads to fracture of the SE. [32] Investigating the combined effects of these various factors requires a model that couples Li electrodeposition and mechanical deformation together.

The objective of the current study is to provide basic models for the deposition and mechanical stability of the interface between Li metal and the SE, under boundary conditions (BCs) that are relevant for practical Li-metal SSBs. We first show that the interfacial stress caused by the stack pressure has a much smaller effect on the Li deposition stability compared to other factors, such as the charge-transfer of $Li^+/Li$ at the interface and the mass-transfer rate of $Li^+$ in the SE. Secondly, we investigate Li electrodeposition and plastic deformation for a wide range of BCs and material properties, including the shape of a surface irregularity, the charge-transfer kinetics at the Li metal/SE interface, the conductivity of bulk SE, the yield strength of Li metal, and the applied stack pressure. Finally, we examine the development of contact loss arising from a finite amount of Li deposition. The increase of surface overpotential as a consequence of contact loss is also discussed. In conclusion, we demonstrate that although ideal stable Li electrodeposition is impossible, the electrochemical and mechanical conditions can be optimized to reduce the electro-deposition instability and ensure mechanical stability of the Li metal/SE interface, thereby allowing control of the overpotential and suppression of dendrite formation.

# Results

## **Model geometry**

The model geometry used for all calculations is shown in Figure 1, where the Li metal anode is on top of a rough SE surface with perfect initial contact. The cell is reduced to a 2D geometry by assuming plane strain and isotropic transport properties in the out-of-plane z-direction. This assumption is reasonable since the surface defects on the SEs are usually long in one dimension compared to the other two dimensions [8], which has been widely employed in similar work. [5,22,23] A good initial contact can be experimentally achieved by applying large stress or heat treatment. [26,33] Here, we mainly focus on the interfacial morphology evolution with continued Li deposition at the interface. We first study the inhomogeneities of current density, overpotential and stress at the Li metal/SE interface caused by the conformal defects of the initial geometry, and then study the evolution of these inhomogeneities when the interface evolves (For example, contact loss may develops if the plastic flow of the Li metal in the confined space is not enough).

In the present work, the Li deposition stability and the mechanical stability are described by the electro-deposition stability factor $\vartheta$ and the contact loss factor $\gamma$ respectively. The deposition stability factor is defined as $\vartheta = i_n(peak)/i_n(valley)$, where $i_n(peak)$ and $i_n(valley)$ are the current densities at the peak and valley of the Li protrusion, respectively, as defined in other work. [22,34] If the current density ($i_n$) is evenly distributed at the interface $S$, the stability factor $\vartheta = 1$, corresponding to homogeneous deposition of metallic Li along the interface $S$. If $\vartheta \gg 1$, Li deposition at the interface $S$ is strongly inhomogeneous. The mechanical stability is defined by the degree of interfacial contact: $\gamma = A_s(t)/A_s(0)$, where $A_s(t)$ is the remaining contact area after $t$ seconds and $A_s(0)$ is the initial contact area. If γ = 1, the initial conformal contact is maintained. When γ < 1, interfacial contact loss develops, which not only causes larger uneven Li deposition but also increases the surface overpotential at interface *S*.

Both the Li electrodeposition and interfacial contact loss are affected by the charge-transfer reactions at interface S (described by the Butler-Volmer relation), mass transfer in the SE (described by the Ohmic relation), and interfacial contact mechanics (described by elastoplastic continuum mechanics). Equations and assumptions associated with the related physical processes are reviewed and discussed in the Method section.

## **Effect of initial surface irregularity on Li deposition stability**

As the presence of surface defects (such as cracks/voids) on the interface *S* is common on the SE surface [7,8], we first study the effect of such defects on Li deposition. We guide ourselves by experimental SE surface roughness data, as for example, determined through atomic force microscopy (AFM) measurements. [35] Several critical parameters are needed to define the surface roughness, including the arithmetic mean length (the average length of defects on a rough surface, $l$) and mean width ($w$), area percent (the ratio of the real surface area to the sampling area), and kurtosis (the shape of defects, with a larger value indicating a sharper defect). For a typical SE material, the mean defect length and width range from 0.01 to 1 $\mu m$, and the kurtosis varies from -1 to 10. [35,36]

Figure 2a shows the model geometry where a small interface irregularity is created at the top of the SE with varying length ($l$), width ($w$), and shape (cosine-shaped, elliptical, or circular). Perfect initial contact between Li metal anode and the SE is assumed and a constant total current density ($i_0 = 0.1 \ mA/cm^2$) is applied at the bottom boundary. Small irregularities (both defect length and width are much smaller than cell size: $\frac{l}{L} \ll 1, \frac{w}{W} \ll 1$) are introduced to ensure that any perturbations of the current density induced by the defect vanish at the boundaries, as indicated by the vertical flow of current density (red arrows) when approaching the boundary $x = \pm 0.5\ \mu m$. The current contour lines in Figure 2a show that current is concentrated near the defect, causing an uneven distribution of the normal current density $i_n$. Figure 2b shows the perturbations affected by defects of typical sizes ($w = 40\ nm, l = 100\ nm$) that are obtained after fine surface polishing of the SE. [8] While the current density near the defect is highly perturbed, the current density $i_n$ at the interface $S$ reaches a constant value ($\sim 0.1\ mA/cm^2$) for $|x| > 0.2\ \mu m$, implying that the perturbation from the defect does not affect the region beyond ten times the defect width. For typical solid electrolyte materials, the contact surface area can be as little as 10% of the total surface area in simple micro/nano polishing. [35,37] This means that most irregularities at Li metal/SE interface will only cause "local" effects to the current density and potential distribution for a well-polished SE surface.

The inset in Figure 2b shows that the peak value of the current density for the three different shaped defects is almost the same ($\sim 0.11\ mA/cm^2$), with the cosine-shaped defect exhibiting a lower magnitude difference ($\vartheta = i_n(peak)/i_n(valley)$) than the elliptical or circular defect. Because the current differences between the three shapes are small, only cosine-shaped defects are considered for further calculation and discussion.

Figure 2c plots the charge-transfer overpotential $\eta$ of three defects with different lengths ($l = 10\ nm, 100\ nm, 200\ nm$) and the same width ($w = 40\ nm$). As the defect length increases, the overpotential becomes more concentrated near the defect and its maximal value grows. For a defect with length of 1 $\mu m$ and width of 8 $nm$ under applied current density $1\ mA/cm^2$, the maximal overpotential reaches $3\ mV$. If unscreened, this large overpotential could generate a stress with magnitude $25\ MPa$ according to the thermodynamic relation: $\sigma = \frac{F}{V_{Li}}\eta$, which is sufficient to crack SE materials with low fracture toughness. [8]

A more systematic study of the geometry effect (defect width $w$ and length $l$) on the stability factor $\vartheta$ is done by calculating stability factors for defects with various widths and lengths, which is shown in Figure 2d. In general, $\vartheta$ decreases when the defect is "shallow and wide" (top right area) and increases when the defect is "deep and narrow" (bottom left area). Therefore, to increase deposition stability, surface engineering methods should be employed to make the SE surface as flat as possible. For example, mechanical polishing can be used to reduce the defect length, and surface corrosion and etching can be applied to increase the defect width or modify the defect shape [36]. Moreover, Figure 2d shows that for a long defect, $\vartheta$ is very sensitive to $w$. Thus, for a defect length $> 10\ \mu m$, a small increase of the defect width can notably homogenize the plating of metallic Li and increase the deposition stability.

The interfacial area-specific resistance can be computed as $ASR_{interface} = \eta/i_n$, where $\eta$ is the overpotential at interface S and $i_n$ is the normal current density at the same point. Notably, $ASR_{interface}$ is dependent of the interfacial contact loss and the exchange current density at the interface. In experiments, the interfacial resistance can be measured through electrochemical impedance spectroscopy (EIS), and the measured value is usually a combined effect of the interfacial contact loss and the interfacial charge transfer, which explains why the measured values can be very different even with the same material. For example, ASR values between 0.2 and $200\ \Omega \cdot cm^2$ have been measured for Li/LLZO/Li cells depending on the applied pressure. [20] To decouple the contact loss resistance and the charge transfer resistance, intimate interfacial contact is needed, which can be achieved either by applying large enough stack pressure (400 $MPa$ stack pressure gives ~0.2 $\Omega \cdot cm^2$ for LLZO [20]), or by smooth enough surface conditions (10 $nm$ surface roughness gives ~0.5 $\Omega \cdot cm^2$ for LiPON [35]). When intimate contact is assumed in our work, the interfacial resistance $ASR_{interface}$ is 0.26 $\Omega \cdot cm^2$, in good agreement with the above experiments. Effects from contact loss will be discussed in the last section.

## **Effect of charge transfer kinetics on deposition stability**

Additional factors that affect the deposition stability are the SE ionic conductivity $\sigma_{Li^+}$ and the exchange current density $i_{exc}$, which controls the $Li^+$ mass-transfer rate at the interface, as shown in Equation 5. Because the formation of a SEI is common between the electrode and SE (such as Li metal with LPS), substantial changes of the ionic conductivity $\sigma_{Li^+}$ ($10^{-3}$-$10^{-1}$ $mS/cm$) [38] and exchange current density $i_{exc}$ (10-$10^3$ $mA/cm^2$) [39,40] are considered in this study.

The effect of ionic conductivity $\sigma_{Li^+}$ and exchange current density $i_{exc}$ on Li deposition is studied with the same model shown in Figure 2a. A constant total current density ($i_0 = 0.1$ $mA/cm^2$) is applied at the bottom boundary. In Figure 3a, the normal current density $i_n$ at interface $S$ is plotted for three different values of $i_{exc}$ with $\sigma_{Li^+} = 0.3$ $mS/cm$. [38] While a small exchange current density $i_{exc}$ leads to a higher overpotential, it can homogenize the current distribution when it becomes the rate limiting step. Figure 3b shows the current density $i_n$ for three different values of ionic conductivity $\sigma_{Li^+}$ with $i_{exc} = 100$ $mA/cm^2$. [39] A larger ionic conductivity $\sigma_{Li^+}$ corresponds to faster $Li^+$ migration in the SE, which also homogenizes the current distribution. Therefore, a smaller exchange current density $i_{exc}$ or larger ionic conductivity $\sigma_{Li^+}$ is preferred for stable Li electrodeposition. The coupled effect of $\sigma_{Li^+}$ and $i_{exc}$ on the deposition stability $\vartheta$ is shown in Figure 3c, with ionic conductivity $\sigma_{Li^+}$ ranging from $3 \times 10^{-4}$ to $3 \times 10^{-1}$ $mS/cm$ and exchange current density $i_{exc}$ ranging from 10 to $10^3$ $mA/cm^2$. It is clear from the figure that both the ionic conductivity and the exchange current density can have a large influence on the stability factor of deposition.

To maintain stable electrodeposition (top left area of Figure 3c) at interface $S$ without dendrite formation and contact loss, a Li metal SSB should possess a SE with high ionic conductivity $\sigma_{Li^+}$ in the bulk and low exchange current density $i_{exc}$ at the interface with Li metal. Notably, a higher $\sigma_{Li^+}$ not only ensures more stable Li deposition but also results in a lower ohmic area-specific-resistance (ASR) of the bulk SE. However, although a lower exchange current density $i_{exc}$ ensures more stable deposition, it may result in a higher charge-transfer ASR at the interface. Figure 3d plots the interfacial charge-transfer ASR as a function of the exchange current density $i_{exc}$. The charge-transfer ASR increases dramatically when $i_{exc}$ is less than 50 $mA/cm^2$. Notably, lots of work has focused on decreasing the interfacial ASR through interface modification [18,41] in order to prevent dendrite formation, which may appear to contradict the conclusion in this work that higher charge-transfer ASR is expected to homogenize the current density. This apparent contradiction occurs because in most experimental measurements, it is hard to separate the local charge-transfer ASR from the total interfacial ASR, which includes contact loss as well as interfacial overpotential, as discussed previously. We believe that in experiments, the decrease of interfacial ASR arises mainly from better interfacial contact. Our conclusion only applies to the part of the ASR that is controlled by the charge transfer.

Conventionally, the formation of a SEI (Figure 1b) between Li metal and the SE is believed to be detrimental to cell performance. One reason is that the ionic conductivity $\sigma_{Li^+}$ within the SEI is usually much lower than that in the bulk SE, leading to a higher ohmic ASR at the SEI and lower Li deposition stability (Figure 3c). However, there are scenarios under which the effectively lowered exchange current density $i_{exc}$ through SEI formation may enhance depositional stability. Therefore, in determining the function of the SEI in a Li metal SSB, both the negative effect on the ASR and the beneficial effect on the depositional stability should be considered.

### **Effect of stack pressure and yield strength on Li metal plasticity and SE fracture**

Unstable Li deposition may cause contact loss at interface *S*. Therefore, the application of a stack pressure $P_0$ is essential during cell cycling to maintain intimate contact. In principle, $P_0$ should be large enough to drive plastic/creep deformation of Li metal to smooth out the inhomogeneities caused by the uneven deposition. A wide range of yield strengths for Li, $\sigma_y$, from 0.6 to 50 $MPa$ has been reported in the literature. [42-44] Although creep behavior of Li metal has been observed in some experiments [42,45,46], it is not included in the current study due to the lack of reliable constitutive measurements. Instead, perfect plasticity without hardening and with different yield strength $\sigma_y$ (ranging from 0.5 to 3 $MPa$) is assumed to reflect the effect of applied current density on the yield strength. [42]

The effects of the stack pressure $P_0$ (ranging from 1 to 5 $MPa$) and the yield strength of Li metal $\sigma_y$ (ranging from 0.5 to 3 $MPa$) on Li deformation are investigated using the model shown in Figure 2a (defect width $w = 80\ nm$ and length $l = 200\ nm$). The existence of surface irregularities will cause local intensification of both the hydrostatic pressure *P* and the von Mises stress $\sigma_v$ at the interface *S*. The latter is the relevant quantity that determines the plastic yield of Li metal. Figure 4a and 4b show the distributions of hydrostatic pressure $P$ and von Mises stress $\sigma_v$ of Li metal along interface $S$ near the Li protrusion for different stack pressures $P_0$ and a yield strength ($\sigma_y = 0.8\ MPa$) before any current is applied. As observed in Figure 4a and 4b, the pressure $P$ increases as the stack pressure $P_0$ increases, reaching a maximal value of $5.8\ MPa$ when the stack pressure is 3 MPa. The interfacial von Mises stress $\sigma_v$ also increases but is bounded by the yield strength $\sigma_y$ ($0.8\ MPa$ in this case). The peak point of the surface irregularity has the largest von Mises stress $\sigma_v$ and therefore reaches plasticity first. When a larger stack pressure $P_0$ is applied, the plastic region of Li metal grows from the peak area.

Figure 4c shows the distribution of the elastic/plastic regions of Li metal under a stack pressure $P_0 = 3\ MPa$ and Li yield strength $\sigma_y = 0.8\ MPa$. The elastic area is colored in light gray (labeled as region III in Figure 4c), and the plastic area is colored in dark blue (labeled as regions I & II). The peak area of the Li protrusion (region I) reaches plasticity due to stress intensification, which is consistent with the results in Figure 4b. It is more difficult for the top area of the Li protrusion (region III) to reach plasticity than the area far away from the protrusion (region II). This result stems from the hydrostatic stress in region III being much higher than the Von-Mises stress because of the confining geometry, consistent with the plastic deformation of a material in a confined space. [47] The degree of plasticity $\beta$ (the ratio of the plastically deformed area to the total area of Li metal) can be used to describe the plastic flow of Li metal. Figure 4d shows the effect of the stack pressure $P_0$ and yield strength $\sigma_y$ on $\beta$, with minimum value 0 (fully elastic) and maximum value 1 (fully plastic).

From Figure 4c, we can see that the location of Li plasticity may play a role in the cell performance an aspect which has not been previously discussed in other studies. In Figure 4c, region I and II reach plasticity while region III only elastically deforms. Note that the maximal elastic deformation for metals is typically very small and for Li metal is ∼0.01%.[46] The lack of plastic flow in region III is destructive because any voids/defects developed in this region may not be filled in by Li plastic flow because of the limited deformation taking place. When a larger stack pressure (such as $P_0 = 3.5 MPa$) is applied, region III will disappear and the Li metal reaches full plasticity ($\beta = 1$). As the stack pressure keeps on increasing, larger pressure intensification will develop near the defect area of the Li metal since the Von-Mises stress is bounded by its yield strength. This stress intensification near the defect area in both the Li metal and the SE may cause the fracture of the SE if the defect is sharp and the stack pressure is large enough.[48]

In Figure 4d, the degree of plasticity $\beta$ increases as the stack pressure $P_0$ increases or the yield strength $\sigma_y$ decreases. Considerable debate exists in the literature as to the yield stress of Li.[42,43,45] To maintain a high enough plasticity (such as $\beta = 0.8$ shown in Figure 4d), the stack pressure $P_0$ must be increased accordingly if the yield strength of Li metal were to be larger than the accepted macroscale value ($0.8\ MPa$). For example, if the yield strength of Li can reach $50\ MPa$ [43], a stack pressure of $150\ MPa$ (extrapolated from the curve $\beta = 0.8$ shown in Figure 4d) is needed to drive enough Li deformation to remove effects of inhomogeneous current deposition. However, although this high stack pressure $P_0$ can be used to promote the Li metal deformation, it can also result in fracture of the SE material. Such SE fracture under cell operation conditions has been observed in experimental studies.[32] A detailed analysis of the fracture mechanisms of solid electrolytes is beyond the scope of this paper. One can assess the effects of stack pressure $P_0$ and the relative Young's moduli of electrolyte and Li metal ($E_r = E_{SE}/E_{Li}$) on fracture through an elementary fracture mechanics model [49], as summarized in Figure 5.

In order to provide direct comparison with experimental studies [48,50], the model built in Figure 5 follows the real defect sizes measured in these studies: defect width $w = 1\ \mu m$ and length $l = 20\ \mu m$. SE fracture will occur near the tip of defect and propagate if the stress intensity factor under mode I loading condition ($K_I$) is larger than the fracture toughness ($K_{IC}$) of the SE material. [49] According to the G-criterion, the stress intensity factor $K_I$ and the strain energy release rate ($G$) around the defect are related by: $K_I = \sqrt{G \cdot \frac{E_{SE}}{1-\nu^2}}$, $\nu$ is the Poisson's ratio of the SE. [49] The strain energy release rate ($G$) can be extracted from the stress state near the defect tip through the use of J-integral. [51] The maximal allowed stack pressure $P_0$ to prevent SE fracture should satisfy the relation: $K_I(P_0, E_r) \leq K_{IC}$. (More details are shown in section 1-d).

Figure 5a shows the stress intensity factor ($K_I$) as a function of the applied stack pressure $P_0$ for two different Young's moduli ratio $E_r$ ($E_r = 2.5$ for LPS and $E_r = 19.2$ for LLZO) when the yield strength of Li metal $\sigma_y = 0.8\ MPa$. Under the same stack pressure, the stress intensity factor of a rigid SE material (larger $E_r$, red line) is smaller than a flexible SE material (smaller $E_r$, black line). Estimating the maximally allowed stack pressure $P_0$ from the black and red lines in Figure 5a using the published fracture toughness of LPS ($0.25\ MPa \cdot m^{1/2}$) and LLZO ($1.25\ MPa \cdot m^{1/2}$) [31,52] gives $160\ MPa$ for LPS and $3 GPa$ for LLZO, respectively. Therefore, the LLZO-type SE has a higher resistance to fracture than LPS-type SE, not only due to the higher fracture toughness but also because of the lower stress intensity factor under the same stack pressure.

Figure 5b shows the stress intensity factor $K_I$ of the SE as a function of the stack pressure $P_0$ for three different yield strengths of Li metal $\sigma_y$ with an LPS-type SE. Under the same stack pressure, the stress intensity factor is smaller if the yielding point of the Li metal is higher. When the stack pressure $P_0$ is less than $4\ MPa$, the stress intensity factor $K_I$ of the SE are almost identical for the three different yield strengths of Li metal. This is because part of the Li metal is still in the elastic regime, leading to similar mechanical responses of the contact pair. However, as the stack pressure $P_0$ rises, the $K_I - P_0$ curves start to deviate from linearity (The smaller the yielding strength, the earlier the deviation). This is due to the transition of Li metal deformation from elastic to plastic. After Li metal reaches full plasticity, a new linear relation between the stress intensity factor $K_I$ and the stack pressure $P_0$ is observed.

In general, results in Figure 5 show that the fracture of the SE is not only determined by its toughness, but also by its stiffness and by the yielding strength of the metal anode. An SE with higher fracture toughness and Young's modulus paired with a metal anode with high yielding strength are preferred in order to increase the fracture resistance of the cell.

## **Coupled effect of Li electrodeposition and Li plasticity on contact loss**

Whether contact loss occurs at an interface is determined by the competition between the irregular Li deposition and the ability to displace Li metal under plastic deformation driven by the stack pressure $P_0$. In this section, both the mechanical and electro-deposition effects on interfacial contact loss are discussed. To quantify the contact loss, the degree of contact $\gamma$ is defined as $\gamma(t) = A_s(t)/A_s(0)$, where $A_s(t)$ is the remaining interfacial contact area after $t$ seconds and $A_s(0)$ is the initial contact area. For the 2D model used in this work, the "contact area" is the contact length of interface $S$.

A similar model as that in Figure 2a is built with defect width $w = 40\ nm$ and length $l = 200\ nm$, and the degree of contact $\gamma$ at interface $S$ is investigated for different values of the exchange current density $i_{exc}$ ($30 - 160\ mA/cm^2$), applied total current density $i_0$ ($0.01 - 0.2\ mA/cm^2$), stack pressure $P_0$ ($1 - 5\ MPa$), and yield strength $\sigma_y$ ($0.8 - 4\ MPa$), with the ionic conductivity $\sigma_{Li^+}$ maintained at $0.3\ mS/cm$. The thickness of Li deposited is calculated from the Faradaic relation $H = \frac{V_{Li}}{F} i_n \Delta t$, where $H$ is the thickness of newly deposited Li and $\Delta t$ is the deposition time. The simulation proceeds by depositing Li metal for a discrete timestep, followed by a mechanical equilibration after each timestep, including elastic and plastic deformation. (A convergence test is done on the selection of the timestep, as described in section SI-6). The process is shown in more detail in the flowchart in Figure SI-1. A total deposition time of $50s$ (with average deposition thickness $\sim 1nm$) is conducted in all calculations with timestep $\Delta t = 10s$.

Figure 6a and Figure 6b show examples of the contact loss at interface $S$ with conditions: applied total current density $i_0 = 0.1\ mA/cm^2$, exchange current density $i_{exc} = 100\ mA/cm^2$, and yield strength of Li metal $\sigma_y = 0.8\ MPa$. When the stack pressure is low (such as $P_0 = 1\ MPa$ in Figure 6a), the deformation of Li metal is not sufficient to fill in the voids developed during the Li deposition. Voids near the edges of the defect develop, preventing further Li deposition in these area as shown by the current density in Figure 6b after the voids develop. Once the SE and Li metal anode are separated in the void area, a sharp discontinuity in the current density $i_n$ is created. Higher pressure is able to reduce the contact loss, and when $P_0$ is $3\ MPa$ no contact loss occurs during the limited time of the simulation.

Figure 6c shows the degree of contact $\gamma$ for different applied total current density $i_0$ and exchange current density $i_{exc}$ for a Li metal yield strength of $\sigma_y = 0.8\ MPa$ and stack pressure $P_0 = 2\ MPa$. It is clear that higher charge-transfer rate and larger current density leads to more contact loss, consistent with our earlier observation that large exchange current density magnifies the inhomogeneity in the current density arising from surface irregularity. For a low charge-transfer rate ($i_{exc} = 30\ mA/cm^2$), a higher charging rate ($i_0$ up to $0.2\ mA/cm^2$) can be tolerated to maintain physical contact ($\gamma = 1$). However, approximately 30% of the contact is lost when the exchange current density $i_{exc}$ increases to $160\ mA/cm^2$ and the applied total current density $i_0$ increases to $0.2\ mA/cm^2$. The contact loss shown in Figure 6c originates from uneven electrodeposition because of the original defect geometry. In general, when the applied current density $i_0$ is too large (x-axis in Figure 6c) or the exchange current density is too large (curves from top to bottom), the deposition becomes more unstable, and more void space develops.

Figure 6d shows the degree of contact $\gamma$ for various stack pressure $P_0$ and values of the Li metal yield strength $\sigma_y$ with applied current density $i_0 = 0.1\ mA/cm^2$ and exchange current density $i_{exc} = 100\ mA/cm^2$. A low yield strength and a large stack pressure both help to prevent contact loss. When Li metal is soft ($\sigma_y = 0.8\ MPa$), a small value of the stack pressure $P_0$ (~$2.5\ MPa$) is sufficient to maintain perfect contact ($\gamma = 1$). However, approximately 6% of the contact will be lost when the yield strength of Li metal is $3\ MPa$ even if the stack pressure $P_0$ increases to $5\ MPa$. Therefore, the stack pressure $P_0$ needs to be well above the yield strength of Li metal $\sigma_y$ to maintain intimate contact near surface irregularities (As shown in Figure 6d, $P_0 > 2.2\ MPa$ is needed if $\sigma_y = 0.8\ MPa$, and $P_0 > 3.6\ MPa$ is needed if $\sigma_y = 1.3\ MPa$). Moreover, the degree of contact $\gamma$ decreases more rapidly as the yield strength $\sigma_y$ increases; such as for $\sigma_y = 1.8\ MPa$, a minimum value of the stack pressure $P_0 = 5\ MPa$ is needed maintain contact, which is close to the fracture limit (6 $MPa$) for LPS fracture (Figure 5a). The contact loss developed in Figure 6d mainly originates from insufficient mechanical deformation. In general, when the applied stack pressure is too low (x-axis in Figure 6d), or the yield strength of Li metal is high (curves from top to bottom), deformation of Li metal is not sufficient to smooth out the uneven deposition.

## Discussion

Unstable Li deposition is believed to be one of the main sources for dendrite formation/propagation and interfacial contact loss. Both effects will amplify current inhomogeneity by either focusing current on a Li domain that is growing into the SE, or by redirecting current from the contact loss area to other areas where now the local current density needs to increase. As the internal stress of irregularly deposited Li accumulates (Figure 4a and Figure 4b), two different mechanisms can be triggered to release this internal stress: a) The deposited lithium is driven into pores or GBs of the SE, where the mechanical strength of the SE is weaker, initiating dendrite propagation; b) If the stack pressure is low enough, the excess internal stress in deposited Li can also be released by pushing the Li metal anode backward, resulting in interfacial contact loss (Figure 6). Since an ideal homogeneous current density distribution (with stability factor $\vartheta = 1$) is likely to be impossible due to the geometric and/or material inhomogeneities, it is more practical to minimize its detrimental consequences through an optimization of the material's structure (such as surface geometry and pore connectivity of SE) and boundary conditions (such as the stack pressure or applied current density).

To increase deposition stability, both the geometry and the electrochemical properties of Li metal and SE are important: a) "Shallow and wide" surface defects are preferred (Figure 2d). Specifically, a decrease in the surface roughness by ten times results in an increase in deposition stability by approximately 20%. b) A higher ionic conductivity of the SE is always preferred for both decreasing the SE bulk resistance and increasing the deposition stability. For example, interfacial deposition can be 25 times more stable (stability factor $\vartheta$ decreases from 30 to 1.2 when the ionic conductivity $\sigma_{Li^+}$ increases from $0.003\ mS/cm$ to $0.3\ mS/cm$ (Figure 3b). c) For a given defect geometry, a relatively slow interfacial charge-transfer kinetics (low exchange current density) is preferred (Figure 3a) though this has to be kept within what is acceptable in terms of total interfacial resistance (Figure 3d). Increasing the ionic conductivity of the SE is clearly the most efficient way to stabilize the electrodeposition, as decreasing the exchange current density (Figure 3a) or preparing a well-polished SE surface (Figure 2d) only decrease $\vartheta$ by around 3 times and 1.5 times respectively. Therefore, screening for solid electrolyte with a high ionic conductivity [53] is important for the performance of SSB, not only for lower SE resistance but also for more stable interfacial deposition. The reason the ionic conductivity is so important is that it effectively acts to screen surface irregularities. Through Ohm's law inhomogeneous current near a defect also corresponds to inhomogeneities in the electric potential. A conductor with high $\sigma_{Li^+}$ can spread this current laterally in the conductor more easily, thereby effectively screening the defect. This can be well observed in Figure 3b where for high enough ionic conductivity $\sigma_{Li^+}$ the effect of the defect in the current density becomes almost invisible. Hence, we want to stress that high ionic conductivity in a solid-state electrolyte is not only important to lower the impedance of the full cell, but also plays an important role in screening the current inhomogeneities from defects near the charge transfer interface. One very different option to mitigate surface irregularity is to throttle the transport at or very near the interface. This is evidenced in our calculations which show that a smaller exchange current leads to more homogeneous deposition and less void formation. While it seems unlikely that much can be done to change the actual charge transfer kinetics between Li metal and the conductor, a varying exchange current density can be mimicked by using a thin buffer layer. If this thin buffer layer has lower conductivity than the SE it will dampen irregular current coming from the SE over some length scale. But a key requirement is that this buffer layer itself has good contact with the Li metal so that even deposition can occur from the buffer layer. If the buffer layer were to have worse contact with the Li anode, then deposition inhomogeneity would actually be aggravated by its application.

To make our discussion more general, a quantity $l_0 = \sigma_{Li^+} \cdot \frac{RT}{i_{exc}F}$ is defined to reflect the combined effect of the bulk (ionic conductivity $\sigma_{Li^+}$) and the surface (exchange current density $i_{exc}$) of the SE on the Li deposition stability. Notably, the second term $\frac{RT}{i_{exc}F}$ equals the charge transfer area-specific resistance $ASR$, when the Butler-Volmer relation (Equation 5) is linearized (when the overpotential $\eta$ is small: $\eta \ll \frac{RT}{F}$). We call $l_0$ (with length unit) the damping length as it quantifies the ability of the SE to dampen deposition instability caused by surface irregularities. Figure 7a shows the deposition stability as a function of the defect length $l$ and the damping length $l_0$. The dark blue, sky blue and yellow colors correspond to deposition stability factor $\vartheta \leq 10$, $10 < \vartheta < 20$ and $\vartheta \geq 20$, respectively. The two horizontal white lines are the damping length $l_0$ of LPS (28 μm, [39]) and LLZO (2 μm, [41]). As previously discussed, the deposition is more stable when defect length decreases (Figure 2c and 2d) and damping length increases (exchange current density decreases in Figure 3a, or ionic conductivity increases in Figure 3b). An SE with large $l_0$ (such as LPS) creates more stable deposition than an SE with small $l_0$ (such as LLZO) when their surface roughness is similar. That is to say, an SE with larger $l_0$ can tolerate worse surface quality. LPS allows a defect length up to 4 μm while LLZO can only tolerate a value less than 1.4 μm if the deposition stability factor $\vartheta$ has to be kept less than 20. Results in Figure 7a make it possible to evaluate the Li deposition stability at the Li metal/SE interface just from the material and geometry information of the SE.

Accumulation of internal stress near surface irregularities (Figure 4a and 4b) provides a driving force for Li deformation. Low stack pressure may result in insufficient Li deformation (Figure 4d) and cause interfacial contact loss (Figure 6a), while large stack pressure may lead to stress intensification (Figure 5) and cause Li infiltration into pores/GBs or even SE fracture if the stress intensity factor is above the limit of the fracture toughness of the SE. We find that in general, the stack pressure required to cause sufficient Li flow and maintain contact is well above the Li yield stress. For example, a minimal value of $2.2\ MPa$ is needed to maintain intimate contact when the Li yield strength is $0.8\ MPa$ (Figure 6d) and a maximal value of $160\ MPa$ is allowed to prevent SE fracture against LPS (Figure 5a). Therefore, a "mechanical stability window" of the stack pressure ($[\ 2.2\ MPa, 160\ MPa\ ]$) is available to prevent both contact loss and SE fracture. Notably, this window would be modified if the Li metal yield strength is larger than $0.8\ MPa$. This is because larger stack pressure is required to maintain intimate contact at higher Li metal yield strength (Figure 6d) while larger stack pressure is allowed for SE fracture at higher Li metal yield strength (Figure 5b). Therefore, a mechanical stability window for the stack pressure must be established for a Li-metal SSB system depending on the mechanical properties of both Li metal and the SE material.

Conclusions drawn above are also applicable to other metal SSB system (such as Sodium and Magnesium) as long as the metal electrode/SE interface is chemically stable. For solid-state battery system with different combination of metal electrodes and SEs, the required mechanical stability window for stack pressure can be very different. From the available material properties of electrodes and SEs in typical SSB systems (as shown in Table SI-1), we can extract the lower and upper limits of the stack pressure from Figure 6d and Figure 5b respectively. A summary of the mechanical stability window for the stack pressure for typical SSB systems is shown in Figure 7b. Smaller stack pressure is needed in Na-metal system than in Li-metal system to maintain intimate contact. This is because Na-metal has a lower yield strength ($\sigma_y = 0.2\ MPa$) than Li-metal ($\sigma_y = 0.8\ MPa$). Oxide-SE systems can withstand much larger stack pressure than sulfide-SE systems because of their much larger moduli and fracture toughness (Table SI-1). Notably, results shown in Figure 7b assume no hardening of metal electrode and no degradation of the mechanical properties of the SE from voids/GBs.

In this work, the continuum theory with a two-dimensional model is employed to investigate the Li electrodeposition stability and mechanical stability at the Li metal/solid electrolyte interface in Li-metal solid-state battery. Through our analysis, we find that the Li deposition can be stabilized by increasing the ionic conductivity of the SE, decreasing the exchange current density of the interfacial charge-transfer reaction, and preparing a well-polished smooth SE surface. We show that LPS can tolerate worse surface quality than LLZO at the same deposition stability because LPS has much larger "damping length" than LLZO (28 μm vs. 2 μm). We also find that a "mechanical stability window" for the applied stack pressure is important for different SSB systems in order to prevent both contact loss and SE fracture. We conclude that cell performance of SSBs can be enhanced by a more comprehensive selection of SE materials, and a more careful control of applied stack pressure.

# Experimental Procedures

## Lead Contact

Further information and requests for resources and reagents should be directed to and will be fulfilled by the Lead Contact, Gerbrand Ceder (gceder@berkeley.edu).

## Materials Availability

This study did not generate new unique reagents.

## Data and Code Availability

All data and code associated with the study have not been deposited in a public repository because , but available from the lead contact upon reasonable request.

## Mechanical equations

The continuous plating on interface $S$ generates stress fields in both the Li metal and the SE. The presence of initial defects creates inhomogeneous stress distribution, which may be sufficient to drive plastic deformation of Li metal or fracture of the SE [21]. Varying yield strength and creep behavior for Li metal have recently been reported at room temperature.

[45,46] A comprehensive approach combining transport with elasticity and plasticity of both the Li metal and SE can be achieved by solving the following equations:

Quasi-static mechanical equilibrium is assumed for both Li metal and the SE:

$$\nabla \cdot \boldsymbol{\sigma} = \boldsymbol{0} \tag{1}$$

Linear elasticity is assumed for the elastic state of both Li metal and the SE:

$$\boldsymbol{\sigma} = \frac{E}{1+\nu}\boldsymbol{\varepsilon} + \frac{\nu E}{(1+\nu)(1-2\nu)} trace(\boldsymbol{\varepsilon})\boldsymbol{I} \tag{2}$$

An elastic/perfect plastic model without hardening is assumed for the Li metal plastic flow, with the Von Mises criterion and associated flow rule (Details in section SI-1b):

$$\Phi(\boldsymbol{\sigma}) \equiv \sqrt{\frac{3}{2}}|dev(\boldsymbol{\sigma})| - \sigma_y = 0, \qquad d\boldsymbol{\varepsilon}^p = d\lambda \frac{\partial \Phi}{\partial \boldsymbol{\sigma}} \tag{3}$$

At boundary $S$: Interfacial contact equilibrium (More details in section SI-1c).

At boundary $S_1$: Neumann BC with constant compressive stress: $\boldsymbol{\sigma}\boldsymbol{n} = -P_0\boldsymbol{n}$.

At boundary $S_0$: Fixed in both x and y direction. At $S_L$ and $S_R$: Fixed in x direction.

Here, $\boldsymbol{\sigma}$ is the stress tensor; $\boldsymbol{\varepsilon}$ and $\boldsymbol{\varepsilon}^p$ are the total strain tensor and the plastic strain tensor, respectively; $E$ and $\nu$ are the Young's modulus and Poisson's ratio of Li metal or the SE, respectively; $\sigma_y$ is the yield strength of Li metal; $\Phi(\boldsymbol{\sigma}) \equiv 0$ defines the surface of stress state at which yield occurs; $\boldsymbol{n}$ is the unit normal direction of the interface $S$ oriented toward the exterior (for example, the normal of the Li metal, $\boldsymbol{n}_{Li}$, points downward and the normal of the SE, $\boldsymbol{n}_{SE}$, points upward in Figure 1); $P_0$ is the magnitude of the externally applied compressive stress (known as "stack pressure" in experiments); and $\nabla \cdot$, $trace(*)$ and $dev(*)$ are the divergence, trace, and deviatoric operators, respectively. $d\lambda$ is the plastic multiplier; the hydrostatic pressure $P$ ($P = -\frac{1}{3}trace(\boldsymbol{\sigma})$), which may affect the current density $i_n$ as discussed later, and the deviatoric stress $\boldsymbol{s}$ ($\boldsymbol{s} = dev(\boldsymbol{\sigma})$), which drives the plastic flow of metallic Li. The von Mises stress $\sigma_v$ of a material point is defined as:

$\sigma_v = \sqrt{\frac{3}{2}\mathbf{s}:\mathbf{s}}$. The stress intensity factor $K_I$ can be calculated from the stress state of the SE according to the G-criterion, which will be used to describe the fracture of the SE (more details in section SI-1d).

Notably, the boundary conditions defined here are different from that in recent work which concluded that the deposition induced stress can grow up to 1GPa and thereby modify the overpotential and current density distribution in a significant way [54]. Klinsmann et al. assume that Li in a sharp crack cannot be pushed out so that very high stresses can build up at the crack front. Whether or not such boundary condition is applicable, may depend on the adhesion strength between lithium metal and the SE, something that is not well understood at this point. In our model, no adhesive forces between Li metal and SE are present, with the Li electrode only constrained by a Neumann boundary condition at $S_1$ with constant compressive stress, representative of the stack pressure $P_0$ typically used in experiments. We note that recently a more thermodynamic mechanism that may limit the pressure at a crack tip has also been proposed, arguing that any substantial pressure build up at the crack tip will redirect the current away from the crack tip. [55]

In our time resolved simulations, mechanical equilibrium (equation 1) is reached at the end of each timestep for which the electrochemical problem is solved because mechanical equilibrium occurs at a speed much faster than the electro-chemical kinetics in the bulk SE and at the interface $S$. [56]

## Electrochemical equations

Since the SE is a single-ion conductor, it cannot accumulate carriers, and the conduction is therefore purely ohmic (More details in section SI-2a):

$$\nabla^2 \phi_{SE} = 0, \qquad \mathbf{i} = -\sigma_{Li^+} \nabla \phi_{SE} \tag{4}$$

At interface $S$, the Butler-Volmer relation [5,6] is employed as the boundary condition:

$$i_{ct} = i_{exc} e^{\frac{(1-\alpha_a)\overline{V}_{Li}\Delta P_{Li}}{RT}} \left( e^{\frac{\alpha_a F}{RT}\eta} - e^{-\frac{\alpha_c F}{RT}\eta} \right) \quad (5)$$

$$i_{ct} = -i_n = -\boldsymbol{i} \cdot \boldsymbol{n}_{SE} \quad (6)$$

At boundary $S_0$, constant current flows into the SE: $\boldsymbol{i} = \boldsymbol{i}_0$.

At boundary $S_L$ and $S_R$, no current flows out of the sample box: $\boldsymbol{i} \cdot \boldsymbol{n} = 0$.

$\phi_{SE}$ is the electric potential in the SE; $\boldsymbol{i}$ is the current density in the SE; $i_{ct}$ is the ion current density across the interface $S$ (positive for anodic reaction); $i_{exc}$ is the exchange current density of the $Li/Li^+$ reaction; $\alpha_a$ and $\alpha_c$ are the anodic and cathodic charge transfer coefficients, respectively; $\eta$ is the surface overpotential incorporating the mechanical effect: $\eta = \phi_{Anode} - \phi_{SE} - V_0 - \frac{\overline{V}_{Li}\Delta P_{Li}}{F}$, where $\phi_{Anode}$ is the applied electric potential at the Li metal anode and $V_0$ is the unstressed equilibrium potential; $\overline{V}_{Li}\Delta P_{Li}$ is the mechanical energy induced by the local hydrostatic pressure at the interface [5,6] (Details in section SI-2b). $i_n$ is the current density at the surface $S$ of the SE normal to the rough interface $S$ ($i_n = \boldsymbol{i} \cdot \boldsymbol{n}_{SE}$).

Equation 4 implies that $Li^+$ ions are homogeneously distributed in both the bulk and on the surface of the SE without the presence of a "space charge layer". Equation 5 is the Butler-Volmer equation that incorporates the effect of interfacial stress, which was first studied by Monroe and Newman [5,6] and extended in other more recent work [34,54]. Equation 6 is the mass continuity equation at the interface $S$ implying that only the $Li^+$ ions current along the normal direction of the surface $S$ are involved in the plating reaction.

In general, there are two mechanisms by which stress modifies the Butler Volmer equation for non-stress conditions; 1) Most importantly, the hydrostatic pressure modifies the equilibrium potential due to a change in free energy of the metal electrode. This term is $\overline{V}_{Li}\Delta P_{Li}$ and is included in our work through the modification of the overpotential $\eta$. 2) It is also possible that the exchange current density ("$i_{exc}$" in the Butler-Volmer relation) varies due to changes in the barrier for ion transfer from the electrolyte and metal electrode. Lacking a direct determination of this contribution it is often related to the changes in free energy of the end states in the electrode and electrolyte. A recent overview of the various assumptions by which this can be done is given by Ganser et al. [57]. In the approximation made by Newman [5] this change in barrier leads to the extra prefactor $\exp\left[\frac{(1-\alpha_a)\overline{V}_{Li}\Delta P_{Li}}{RT}\right]$ in Equation 5. Because under most of these assumptions the change in $i_{exc}$ is very small for the pressure values seen in this work, we have neglected this effect pressure on exchange current density (but retain the effect of pressure on the overpotential). For a typical value of stack pressure when sulfide-type electrolytes are used (For example, 1 to 5 MPa), $\Delta P_{Li}$ developed in the Li metal near interface $S$ is no more than ~10 MPa (even considering the stress intensification at the peak of defects, as shown in the Results section). At room temperature, the term corresponding to the mechanical effect on the exchange current, $\exp\left[\frac{(1-\alpha_a)\overline{V}_{Li}\Delta P_{Li}}{RT}\right]$, is smaller than 1.02, which implies a very small change of exchange current density at interface $S$ due to hydrostatic pressure.

Notably, the hydrostatic pressure ($P = -\frac{1}{3}trace(\boldsymbol{\sigma})$) was used in this work to quantify the effect of stress on the overpotential ($\overline{V}_{Li}\Delta P_{Li}$), while the normal stress component ($\sigma_n = (\boldsymbol{\sigma n}_{Li})\cdot\boldsymbol{n}_{Li}$) has also been used in literature. [54] We show in section SI-4 that the magnitude of these two components are almost the same ( $P \approx \sigma_n$ ) since Li metal behaves similar to a fluid. [57] A detailed study of the effect of hydrostatic pressure on the surface overpotential is shown in section SI-5. Within the range of practically used stack pressure (1 ~ 5 MPa), the mechanical effect is very small as compared to the effect of inhomogeneities from surface irregularities and transport kinetics. However, this effect can be significant if the stack pressure reaches ~1GPa, as studied in other work. [54,55]

All the partial differential equations (PDEs) are summarized in section SI-3a. An in-house code based on the finite element method is used to solve these nonlinear PDEs. A flow chart of the code is presented in Figure SI-1. Default values of parameters used in this work are from reported experimental measurements of reference papers listed in the last column of Table SI-1. The thickness of both Li metal and the SE are set to be 10 $\mu m$ to ensure the inhomogeneities of stress and current become small when reaching the top and bottom boundaries of the cell. Notably, the thickness of Li metal will affect the behavior of the Li metal anode during cell cycling (such as pop-up delamination [58]).

## Supplementary information

Supplemental information can be found with the article online.

## Acknowledgements

This work was supported by the Office of Energy Efficiency and Renewable Energy of the U.S. Department of Energy under Contract No. 1384-1778. This research used the Lawrencium computational cluster resource provided by the IT Division at Lawrence Berkeley National Laboratory (supported by the Director, Office of Science, Office of Basic Energy Sciences, of the U.S. Department of Energy under Contract No. DE-AC02-05CH11231) and the Savio computational cluster resource provided by the Berkeley Research Computing program at the University of California, Berkeley (supported by the UC Berkeley Chancellor, Vice Chancellor for Research, and Chief Information Officer).

# Author contributions

Q.T. planned the project with G.C.; Q.T. derived all equations, programmed all codes and calculated all data with the help from L.B. and T.S.; The manuscript was written by Q.T. and was revised by L.B., T.S. and G.C. with the help of the other authors. All authors contributed to discussions.

# Declaration of interests

The authors declare no competing interests.

# Figure/table titles and legends

**Figure 1**. (**A**) Schematic model of a half SSB cell consisting of a Li metal anode $\Omega_1$ and solid electrolyte $\Omega_0$. The roughness at the interface $S$ is approximated by the periodic protrusion defined in (c). The surfaces $S_0$ and $S_1$ are flat. The left and right boundaries of Li metal and SE are labeled as $S_L$ and $S_R$, respectively. A stack pressure $P_0$ is applied to the ends of the cell ($S_0$ and $S_1$). Li ions are extracted from the cathode (not shown) and migrate from $S_0$ toward $S$ under galvanostatic conditions: $I_0 = i_0 \cdot A_0 = \int i_n \, dS$, where $i_0$ is the applied current density on $S_0$ with constant area $A_0$, and $i_n$ is the normal current density on $S$ ($i_n = \boldsymbol{i} \cdot \boldsymbol{n}$, where $\boldsymbol{i}$ is the current density in the SE and $\boldsymbol{n}$ is the normal direction of the interface $S$). Upon reaching $S$, these ions are plated onto $\Omega_1$ via the reaction $Li^+ + e^- \leftrightarrow Li$. (**B**) *Li* deposition on the interface $S$ with normal current density $i_n$. A thin layer of SEI above $S$ is shown in yellow. **c).** Geometry of the contact pair of Li metal protrusion and SE defect on the interface $S$, with shape $y(x,t)$, width $2w$, and length $l$.

**Figure 2**. (**A**) Current density distribution in SE for a cosine-shaped defect with defect width $w = 40 \, nm$ and defect length $l = 100 \, nm$; the magnitude is represented by the contour line, and the direction is represented by the red arrows. (**B**) Distribution of normal current density $i_n$ at interface $S$ within range $|x| < 0.2 \, \mu m$ for different shapes with $w = 40 \, nm$ and $l = 100 \, nm$. The inset shows $i_n$ within the range $|x| < 0.05 \, \mu m$. (**C**) Surface overpotential $\eta$ near a defect with different length $l$ but same width $w = 40 \, nm$. (**D**) Stability factor $\vartheta$ (contour surface) as a function of length $l$ and width $w$ for cosine-shaped defect.

**Figure 3**. Distribution of normal current density $i_n$ at interface $S$ near the defect with width $w = 40\ nm$ and $l = 100\ nm$: **(A)** varying exchange current density $i_{exc}$ while keeping ionic conductivity $\sigma_{Li^+}$ constant ($0.3\ mS/cm$) and **(B)** varying $\sigma_{Li^+}$ while keeping $i_{exc}$ constant ($100\ mA/cm^2$). **(C)** Stability factor $\vartheta$ (color map) as a function of the exchange current density $i_{exc}$ and $log\ (\sigma_{Li^+})$. **(D)** Charge-transfer ASR as a function of exchange current density $i_{exc}$. All data at applied current density of 0.1mA/cm².

**Figure 4**. Hydrostatic pressure **(A)** and von Mises stress **(B)** of Li metal along interface $S$ near the defect for a Li metal yield strength $\sigma_y = 0.8\ MPa$. **(C)** Elastic/plastic region of Li metal in contact with the SE (light gray indicates the elastic region, and dark blue indicates the plastic region), with the stack pressure $P_0 = 3\ MPa$ and the yield strength $\sigma_y = 0.8\ MPa$. **(D)** Degree of plasticity $\beta$ of Li metal as a function of the yield strength $\sigma_y$ (varying from 0.5 to 2.5 $MPa$), and the stack pressure $P_0$ (varying from 0.5 to 4.5 $MPa$). The width and length of the Li protrusion (SE defect) are $w = 80\ nm$ and $l = 200\ nm$, respectively.

**Figure 5**. **(A)** Stress intensity factor $K_I$ of SE near the defect tip as a function of the applied stack pressure $P_0$ at two different relative Young's moduli $E_r$ ($E_r = E_{SE}/E_{Li}$, the ratio of Young's moduli between SE and Li metal), with the yield strength $\sigma_y$ fixed at $0.8\ MPa$. **(B)** Stress intensity factor $K_I$ as a function of the applied stack pressure $P_0$ at three different values of the yield strength of Li metal $\sigma_y$, with the relative Young's moduli $E_{Li}$ fixed at 2.5 (for LPS-type SE).

**Figure 6**. (**A**) Loss of interfacial contact after a finite amount of Li deposition with applied total current density $i_0 = 0.1\ mA/cm^2$, stack pressure $P_0 = 1\ MPa$, and Li metal yield strength $\sigma_y = 0.8\ MPa$. (**B**) Current distribution near defect for the conditions in (a) but with three different stack pressure $P_0$. (**C**) Degree of contact for different values of exchange current density $i_{exc}$ and applied total current density $i_0$. (**D**) Degree of contact for different values of Li metal yield strength $\sigma_y$ and stack pressure $P_0$.

**Figure 7**. (**A**) the deposition stability $\vartheta$ as a function of defect length $l$ and damping length $l_0$. The two black dash lines ($\vartheta = 10$ and $\vartheta = 20$) separate the deposition stability into three regions: dark blue ($\vartheta \leq 10$), light blue ($10 < \vartheta < 20$), and yellow ($\vartheta \geq 20$). The two white lines represents the values of damping length for LPS and LLZO respectively. (**B**) The mechanical stability window of the stack pressure for different SSB systems. The lower limit in each window is the minimal value of stack pressure needed to maintain intimate contact of the metal electrode and SE. The upper limit is the maximal value allowed to prevent SE fracture.

Table 1. Critical parameters used in this work

| | Name | Symbol | Unit | Value | Ref. |
|---|---|---|---|---|---|
| **Mechanical Parameters** | Young's modulus of Li metal | $E_{Li}$ | $GPa$ | 7.8 | 46 |
| | Poisson's ratio of Li metal | $\nu_{Li}$ | 1 | 0.38 | 46 |
| | Density of Li metal | $\rho_{Li}$ | $kg/m^3$ | 534 | 46 |
| | Yield strength of Li metal | $\sigma_0$ | $MPa$ | 0.8 | 46 |
| | Young's modulus of SE | $E_{SE}$ | $GPa$ | 19.5 | 31 |
| | Density of SE | $\rho_{SE}$ | $kg/m^3$ | 1880 | 31 |
| | Poisson's ratio of SE | $\nu_{SE}$ | 1 | 0.36 | 31 |
| | Fracture toughness of SE | $K_{IC}$ | $MPa \cdot m^{1/2}$ | 0.25 | 31 |
| **Electro chemical Parameters** | Anodic/cathodic transfer coefficients | $\alpha_a, \alpha_c$ | 1 | 0.5 | 34 |
| | Exchange current density | $i_{exc}$ | $mA/cm^2$ | 100 | 39 |
| | Li⁺ conductivity in SE | $\sigma_{Li^+}$ | $mS/cm$ | 0.3 | 38 |
| | Applied current density | $i_0$ | $mA/cm^2$ | 0.1 | 8 |